\documentclass[12pt]{article}%
\usepackage{amsmath}%
\usepackage{amsfonts}%
\usepackage{amssymb}%
\usepackage{graphicx}

\begin{document}

\title{Statistical models of mixtures with a biaxial nematic phase}
\author{E. do Carmo, D. B. Liarte, and S. R. Salinas
\and Instituto de F\'{\i}sica, Universidade de S\~{a}o Paulo
\and Caixa Postal 66318, CEP 05314-970, S\~{a}o Paulo, Brazil}
\date{14 January 2009}
\maketitle

\begin{abstract}
We consider a simple Maier-Saupe statistical model with the inclusion of
disorder degrees of freedom to mimic the phase diagram of a mixture of
rod-like and disc-like molecules. A quenched distribution of shapes leads to
the existence of a stable biaxial nematic phase, in qualitative agreement with
experimental findings for some ternary lyotropic liquid mixtures. An annealed
distribution, however, which is more adequate to liquid mixtures, precludes
the stability of this biaxial phase. We then use a two-temperature formalism,
and assume a separation of relaxation times, to show that a partial degree of
annealing is already sufficient to stabilize a biaxial nematic structure.

\end{abstract}

Quenched and annealed degrees of freedom of statistical systems are known to
produce phase diagrams with a number of distinct features \cite{BinderYoung}.
The ferromagnetic site-diluted Ising model provides an example of a continuous
transition, in the quenched case, which turns into a first-order boundary
beyond a certain tricritical point, if we consider thermalized site dilution
\cite{Wortis}. Disordered degrees of freedom in solid compounds, as random
magnets and spin-glasses, are examples of quenched disorder, which lead to
well-known problems related to averages of sets of disordered free energies.
In liquid systems, however, relaxation times are shorter, and the simpler
problems of annealed disorder are more relevant from the physical perspective.
In this paper, we show that distinctions between quenched and annealed degrees
of freedom are particularly relevant in statistical models of mixtures, which
have been used to account for the elusive biaxial nematic phase of liquid
crystals \cite{DeGennes}.

Uniaxial nematic phases, with the director along a single axis, have been
fully characterized in the phase diagrams of a large number of thermotropic as
well as lyotropic liquid crystals \cite{DeGennes}\cite{Figueiredo}. The
existence of a biaxial nematic structure, however, has been subjected to some
debate \cite{Luckhurst}. A biaxial nematic phase was predicted by calculations
for different lattice models \cite{Alben}, and has been found by Yu and Saupe
\cite{YuSaupe} in the phase diagram of a ternary lyotropic mixture. In the
ordered region of this mixture, in a phase diagram in terms of temperature and
concentration of one of the compounds, there is a nematic biaxial structure
bounded by two distinct uniaxial nematic structures, loosely associated with
either prolate (cylinder-like) or oblate (disc-like) molecular aggregates.
Transitions between the ordered uniaxial nematic and the disordered phases are
discontinuous, in agreement with the Maier-Saupe approach, and transitions
between uniaxial and biaxial structures are continuous, with a critical line
that is supposed to end at a Landau multicritical point. More recent
experimental work has indicated the existence of nematic biaxial structures in
a certain number of new thermotropic compounds, formed by anisotropic
banana-shaped molecules \cite{Luckhurst}.

Although earlier theoretical calculations indicated the existence of a stable
biaxial nematic structure \cite{Alben}, which is also supported by a Landau-de
Gennes expansion \cite{Longa}, mean- field calculations by Palffy-Muhoray and
collaborators \cite{Palffy}, using the Maier-Saupe interactions to consider a
mixture of cylinders and discs, precluded the stability of an intermediate
biaxial structure, except under some special circumstances. A few years ago
this problem was reanalyzed, in terms of a schematic discrete version of the
Maier-Saupe model, in a paper by Henriques and Henriques \cite{Henriques}, who
pointed out the existence and stability of a biaxial nematic phase, bordered
by two critical lines meeting at a Landau multicritical point \cite{Longa}, in
close contact with the experimental findings of Yu and Saupe \cite{YuSaupe}.
The calculations of Henriques and Henriques, however, which can be carried out
for any distribution of molecular shapes, implicitly assumed a quenched
polymorphism, which may not be adequate for these liquid crystalline systems.

Given the scarcity of experimental data on these biaxial phases, the
apparently conflicting theoretical results \cite{Alben}\cite{Photinos}, and a
few recent and not so conclusive simulations \cite{Berardi}\cite{Cuetos},
there is still room for revisiting the statistical problem of a mixture of
cylinders and discs. Along the lines of the work of Henriques and Henriques
\cite{Henriques}, we then perform standard statistical-mechanics calculations
for a simple discrete version of the Maier-Saupe model, which we call the
basic model, with the inclusion of a binary distribution of shapes to mimic a
mixture of prolate and oblate molecules (cylinders and discs). We draw clear
distinctions between quenched and annealed distributions of shapes. In the
quenched case, we show the existence of a stable biaxial phase in the
temperature-concentration phase diagram, in qualitative agreement with
experiments, and in accordance with the work of Henriques and Henriques. In
the annealed case, however, the same model system leads to a unstable biaxial
phase, in connection with the earlier calculations of Palffy-Muhoray and
collaborators, which provides an example of the distinctions between the
effects of quenched and annealed disorder degrees of freedom.

Although the results for the quenched distribution seem to be in agreement
with the experimental phase diagrams, relaxation times in both liquid and
liquid crystalline systems are relatively short, and the physics of disordered
couplings in these liquid systems should be better represented by thermalized
variables. Even in the annealed situation, however, we should make an attempt
to account for a certain degree of separation between relaxation times. We
then resorted to a two-temperature formalism \cite{Nieuwenhuizen}, with
orientational degrees of freedom and disorder variables coupled to different
heat reservoirs, at temperatures $T$ and $T_{\lambda}=nT$. In this application
of the formalism, we show that a small difference of temperatures is already
enough to change the phase diagram of the fully annealed case, and to
stabilize the biaxial nematic structure, which probably provides an
explanation for the elusive nature of this phase behavior.

In analogy to the Curie-Weiss model of ferromagnetism, the Maier-Saupe model
is given by the energy%
\begin{equation}
\mathcal{H}=-\frac{A}{N}\sum_{1\leq i<j\leq N}\sum_{\mu,\nu=1,2,3}\lambda
_{i}\lambda_{j}S_{i}^{\mu\nu}S_{j}^{\mu\nu},
\end{equation}
where $A$ is a coupling constant, the orientational degrees of freedom are
given by%
\begin{equation}
S_{i}^{\mu\nu}=\frac{1}{2}\left(  3n_{i}^{\mu}n_{i}^{\nu}-\delta_{\mu\nu
}\right)  ,
\end{equation}
where $\left\vert \overrightarrow{n}_{i}\right\vert =1$ for $i=1$, $2$, $...$
$N$, and $\left\{  \lambda_{i}\right\}  $ is a set of (disordered) couplings
representing either prolate ($\lambda_{i}=+1$) or oblate ($\lambda_{i}=-1$)
molecular groups. This model is further simplified if we suppose that the
director $\overrightarrow{n}_{i}$ assumes six values only, along the Cartesian
axes, $\overrightarrow{n}_{i}=\left(  \pm1,0,0\right)  $, $\left(
0,\pm1,0\right)  $, and $\left(  0,0,\pm1\right)  $, according to an early
suggestion of Zwanzig.

In the quenched case, $\left\{  \lambda_{i}\right\}  $ is a set of
independent, identical, and identically distributed random variables,
associated with a probability distribution $p\left(  \lambda_{i}\right)  $.
For a given configuration $\left\{  \lambda_{i}\right\}  $, the partition
function is written as%
\begin{equation}
Z_{q}\left\{  \lambda_{i}\right\}  =\sum_{\left\{  \overrightarrow{n}%
_{i}\right\}  }\exp\left[  \frac{\beta}{N}\sum_{1\leq i<j\leq N}\sum_{\mu
,\nu=1,2,3}\lambda_{i}\lambda_{j}S_{i}^{\mu\nu}S_{j}^{\mu\nu}\right]  ,
\end{equation}
where $\beta$ is the inverse of a dimensionless temperature, and we choose
$A=1$. We now use standard Gaussian identities to write the partition function
as an integration over the independent terms of a symmetric tensor
$\mathbf{Q}$. In the thermodynamic limit, taking into account the
self-averaging properties of this problem, we can write
\begin{equation}
Z_{q}=\int\left[  d\mathbf{Q}\right]  \exp\left[  -\beta Nf_{q}\right]  ,
\end{equation}
where%
\begin{equation}
f_{q}=\frac{1}{2}\sum_{\mu}Q_{\mu\mu}^{2}+\frac{1}{2}\left\langle \lambda
_{i}\right\rangle \sum_{\mu}Q_{\mu\mu}-\frac{1}{\beta}\left\langle \ln\left[
2%
{\displaystyle\sum\limits_{\mu}}
\exp\left(  \frac{3}{2}\beta\lambda_{i}Q_{\mu\mu}\right)  \right]
\right\rangle .
\end{equation}
The equations of minima, $\partial g/\partial Q_{\mu\mu}=0$, lead to the usual
traceless tensor order parameter of the Maier-Saupe approach. We then
introduce the standard parametrization $Q_{xx}=-\left(  S+\eta\right)  /2$,
$Q_{yy}=-\left(  S-\eta\right)  /2$, and $Q_{zz}=S$, and consider a
double-delta probability distribution, $p\left(  \lambda_{i}\right)
=c\delta\left(  \lambda_{i}-1\right)  +\left(  1-c\right)  \delta\left(
\lambda_{i}+1\right)  $, where the parameter $c$ represents the concentration
of prolate ($\lambda=+1$) molecules. We then have%
\[
f_{q}=\frac{1}{4}\left(  3S^{2}+\eta^{2}\right)  -\frac{1}{\beta}\ln2-\frac
{c}{\beta}\ln\left[  2\exp\left(  -\frac{3}{4}\beta S\right)  \cosh\left(
\frac{3}{4}\beta\eta\right)  +\exp\left(  \frac{3}{2}\beta S\right)  \right]
-
\]%
\begin{equation}
-\frac{\left(  1-c\right)  }{\beta}\ln\left[  2\exp\left(  \frac{3}{4}\beta
S\right)  \cosh\left(  \frac{3}{4}\beta\eta\right)  +\exp\left(  -\frac{3}%
{2}\beta S\right)  \right]  ,
\end{equation}
which leads to the phase diagram of Fig. 1, with thermodynamically stable
uniaxial ($S\neq0$; $\eta=0$) and biaxial ($S\neq0$; $\eta\neq0$) nematic
structures, and two critical lines meeting at the Landau multicritical point,
$\beta=4/3$ and $c=1/2$.

In order to make contact with the standard Landau-de Gennes phenomenology, we
expand the free energy $f_{q}$ in the neighborhood of the Landau multicritical
point, in terms of the quadratic and cubic, $I_{2}=\operatorname{Tr}%
\mathbf{Q}^{2}$ and $I_{3}=\operatorname{Tr}\mathbf{Q}^{3}$, invariants of the
tensor order parameter,
\[
f_{q}=-\frac{3}{4}\ln6+\frac{1}{2}\left(  1-\frac{3}{4}\beta\right)
I_{2}-\frac{1}{3}\left(  2c-1\right)  I_{3}%
\]%
\begin{equation}
+\frac{1}{12}I_{2}^{2}+\frac{1}{6}\left(  2c-1\right)  I_{2}I_{3}+\frac{1}%
{15}I_{3}^{2}+.....\label{l1}%
\end{equation}
According to published analyses \cite{Longa} of this phenomenological
expansion, a positive coefficient of $I_{3}^{2}\ $is indeed enough to
stabilize the biaxial phase.%

\begin{figure}
[ptb]
\begin{center}
\includegraphics[
height=1.9873in,
width=2.8253in
]%
{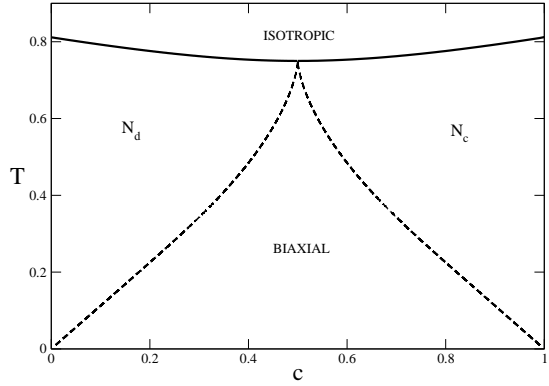}%
\caption{Temperature-concentration phase diagram for a quenched distribution
of shapes. First-order transitions between the isotropic region and the
uniaxial nematic structures (N$_{c}$ and N$_{d}$) are indicated by the solid
line. Dashed lines indicate second-order transitions.}%
\end{center}
\end{figure}

We now turn to the annealed case, which is associated with the canonical
partition function%
\begin{equation}
Z_{a}=%
{\displaystyle\sum\nolimits_{\left\{  \lambda_{i}\right\}  }^{\prime}}
\sum_{\left\{  \overrightarrow{n}_{i}\right\}  }\exp\left[  \frac{\beta}%
{N}\sum_{1\leq i<j\leq N}\sum_{\mu,\nu=1,2,3}\lambda_{i}\lambda_{j}S_{i}%
^{\mu\nu}S_{j}^{\mu\nu}\right]  ,
\end{equation}
where the sum over over $\left\{  \lambda_{i}\right\}  $ is restricted by the
fixed concentrations of the molecular types. As $\lambda=+1$ corresponds to a
prolate and $\lambda=-1$ to an oblate molecule, we have $N_{p}-N_{o}%
=\sum_{i=1}^{N}\lambda_{i}$, and $N_{o}=N-N_{p}$, where $N_{p}$ ($N_{o}$) is
the number of prolate (oblate) molecules, and $N$ is the total number of
molecules. We then introduce a chemical potential and change to the
grand-canonical ensemble,%
\[
\Xi_{a}=%
{\displaystyle\sum\limits_{N_{p}=0}^{N}}
\exp\left(  \beta\mu N_{p}\right)  Z_{a}=
\]%
\begin{equation}
=%
{\displaystyle\sum\limits_{\left\{  \lambda_{i}\right\}  }}
\sum_{\left\{  \overrightarrow{n}_{i}\right\}  }\exp\left[  \beta\mu\frac
{1}{2}\left(  \sum_{i=1}^{N}\lambda_{i}+N\right)  +\frac{\beta}{N}\sum_{1\leq
i<j\leq N}\sum_{\mu,\nu=1,2,3}\lambda_{i}\lambda_{j}S_{i}^{\mu\nu}S_{j}%
^{\mu\nu}\right]  .
\end{equation}
For the same basic model, with six choices of the unit vectors, we have
\begin{equation}
\Xi_{a}=\int\left[  d\mathbf{Q}\right]  \exp\left[  -\beta N\phi_{a}\right]  ,
\end{equation}
with the grand potential per particle%
\[
\phi_{a}=\frac{1}{2}\sum_{\mu}Q_{\mu\mu}^{2}-\frac{1}{2}\mu-\frac{1}{\beta}%
\ln2-
\]%
\begin{equation}
-\frac{1}{\beta}\ln\sum_{\lambda=\pm1}\left[  \exp\left(  \frac{1}{2}%
\lambda\beta\mu-\frac{1}{2}\lambda\beta\sum_{\mu}Q_{\mu\mu}\right)  \sum_{\mu
}\exp\left(  \frac{3}{2}\lambda\beta Q_{\mu\mu}\right)  \right]  ,
\end{equation}
which can also be written as
\[
\phi=\frac{1}{4}\left(  3S^{2}+\eta^{2}\right)  -\frac{1}{2}\mu-\frac{2}%
{\beta}\ln2-
\]%
\begin{equation}
-\frac{1}{\beta}\ln\left[  2\cosh\left(  \frac{3}{4}\beta\eta\right)
\cosh\left(  \frac{1}{2}\beta\mu-\frac{3}{4}\beta S\right)  +\cosh\left(
\frac{1}{2}\beta\mu+\frac{3}{4}\beta S\right)  \right]  .
\end{equation}
Given the thermodynamic fields $T$ and $\mu$, we show that a biaxial solution
exists but that it is thermodynamically unstable, in agreement with the
calculations of Palffy-Muhoray and collaborators. In the $T-\mu$ phase diagram
there is just a first-order boundary, at $\mu=0$, for $T=1/\beta\leq3/4$,
between two distinct uniaxial structures. In this fully annealed situation, we
can also draw the corresponding temperature-concentration phase diagram, as
shown in Fig. 2, in which we sketch the tie lines of coexistence of the two
uniaxial nematic phases.

In order to make contact with the Landau-de Gennes phenomenology, we can also
write an expansion in the neighborhood of the Landau multicritical point,
$\beta=4/3$ and $\mu=0$,%

\[
\phi_{a}=-\frac{3}{4}\ln12+\frac{1}{2}\left[  1-\frac{3}{4}\beta\right]
I_{2}-\frac{2}{9}\mu I_{3}+
\]%
\begin{equation}
+\frac{1}{12}I_{2}^{2}-\frac{1}{18}I_{2}I_{3}-\frac{1}{135}I_{3}%
^{2}+.....\label{l2}%
\end{equation}
In contrast to Eq. (\ref{l1}), the coefficient of $I_{3}^{2}$ is negative,
which now precludes the stability of a biaxial nematic structure \cite{Longa}.%

\begin{figure}
[ptb]
\begin{center}
\includegraphics[
height=2.6109in,
width=3.7144in
]%
{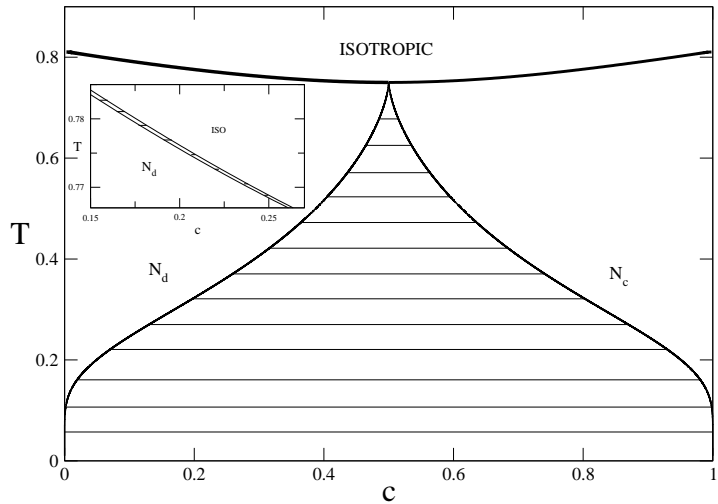}%
\caption{Temperature-concentration phase diagram in the fully annealead case.
The horizontal tie lines indicate the coexistence between two distinct
uniaxial nematic phases. In the inset, we show an amplification of a section
of the (narrow) coexistence region between uniaxial nematic and isotropic
phases. There is no stable biaxial nematic structure.}%
\label{fugure 2}%
\end{center}
\end{figure}

If we use standard techniques of statistical physics, it is straightforward to
show that the Helmholtz free energy $f_{a}$ of the annealed version of the
basic Maier-Saupe model differs from the corresponding free energy $f_{q} $ of
the quenched version by a term corresponding to the entropy of mixing,%
\begin{equation}
f_{a}=f_{q}+\frac{1}{\beta}\left[  c\ln c+\left(  1-c\right)  \ln\left(
1-c\right)  \right]  .
\end{equation}
This is indeed a quite general result, which we have been able to show for a
class of mean-field self-averaging disorder variables, with the insertion of
the appropriate form of the entropy of mixing. It is interesting to point out
a resemblance with a derivation by Mazo of a quenched Helmholtz free energy
that includes, in addition to the expectation of the logarithm of the
partition function, a usually inaccessible term of entropy \cite{Mazo}. We
remark that quenched situations are far from true thermodynamic equilibrium,
and that in the quenched case we do not have access to the entropy of mixing.
Of course, the thermodynamic analysis of the Helmholtz free energy $f_{a}$,
which should be a properly convex function of the density $c$, leads to the
same results of the analysis of the grand potential $\phi_{a}$, written in
terms of the thermodynamic fields, $T$ and $\mu$.

We now search for a stable biaxial phase in a situation of partial annealing,
which may be represented by the introduction of two heat baths, at different
temperatures, associated with the relaxation times of the orientational
(quicker) and disorder (slower) degrees of freedom \cite{Nieuwenhuizen}. Given
a configuration $\lambda$ of the slower disorder variables, we can
schematically write the probability of occurrence of a configuration $\sigma$
of the orientational variables,%
\begin{equation}
P\left(  \sigma\left\vert \lambda\right.  \right)  =\frac{1}{Z_{\sigma}}%
\exp\left[  -\beta\mathcal{H}\left(  \sigma,\lambda\right)  \right]  ,
\end{equation}
where $T=1/\beta$ is the temperature of a heat bath, and
\begin{equation}
Z_{\sigma}=Z_{\sigma}\left(  \lambda\right)  =\sum_{\sigma}\exp\left[
-\beta\mathcal{H}\left(  \sigma,\lambda\right)  \right]  .
\end{equation}
The time evolution of $\lambda_{i}$ is given by a Langevin equation,%
\begin{equation}
\Gamma\frac{\partial\lambda_{i}}{\partial t}=-z\left(  t\right)  \lambda
_{i}-\frac{\partial\mathcal{H}}{\partial\lambda_{i}}+\eta_{i}\left(  t\right)
,
\end{equation}
where $z\left(  t\right)  $ is a multiplier associated with the chemical
potential, and
\begin{equation}
\left\langle \eta_{i}\left(  t\right)  \eta_{j}\left(  t^{\prime}\right)
\right\rangle =2\Gamma T_{\lambda}\delta_{ij}\delta\left(  t%
\acute{}%
-t^{\prime}\right)  ,
\end{equation}
where we have introduced the temperature $T_{\lambda}$ of a second heat bath.
With the assumption of quick and slow time scales, it is reasonable to replace
$\partial\mathcal{H}/\partial\lambda_{i}$ by its average value,%
\begin{equation}
\frac{\partial\mathcal{H}}{\partial\lambda_{i}}\implies\left\langle
\frac{\partial\mathcal{H}}{\partial\lambda_{i}}\right\rangle _{\sigma}%
=\frac{\partial\mathcal{H}_{eff}}{\partial\lambda_{i}},
\end{equation}
where%
\begin{equation}
\mathcal{H}_{eff}=\mathcal{H}_{eff}\left(  \lambda\right)  =-k_{B}T\ln
Tr_{\sigma}\exp\left[  -\beta\mathcal{H}\left(  \sigma,\lambda\right)
\right]  .
\end{equation}
We then assume that the probability of a configuration $\left\{  \lambda
_{i}\right\}  $ is given by the grand-canonical expression%
\begin{equation}
P\left(  \lambda\right)  =\frac{1}{\Xi\left(  \beta_{\lambda},\beta
,N,\mu\right)  }\exp\left[  \beta_{\lambda}\mu N_{p}-\beta_{\lambda
}\mathcal{H}_{eff}\right]  ,
\end{equation}
where
\begin{equation}
\Xi\left(  \beta_{\lambda},\beta,N,\mu\right)  =\int\left[  d\lambda\right]
\left\{  \sum_{\left\{  \sigma\right\}  }\exp\left[  -\beta\mathcal{H}\left(
\sigma,\lambda\right)  +\frac{\beta\mu}{2}\left(
{\textstyle\sum\limits_{i}}
\lambda_{i}+N\right)  \right]  \right\}  ^{\mathbf{n}},
\end{equation}
with $n=T/T_{\lambda}$, which resembles the number of replicas in spin-glass
problems. The application of this formalism to the basic Maier-Saupe model
leads to a grand potential $\phi$, with the following expansion about the
Landau multicritical point, $\beta=4/3$ and $\mu=0$,%

\[
\phi=-\frac{3}{4}\ln6-\frac{3}{4n}\ln2+\frac{1}{2}\left[  1-\frac{3}{4}%
\beta\right]  I_{2}-\frac{2}{9}n\mu I_{3}+
\]%
\begin{equation}
+\frac{1}{12}I_{2}^{2}+\frac{2}{27}\left[  \frac{9}{10}-n\operatorname{sech}%
^{2}\left(  \frac{2}{3}n\mu\right)  \right]  I_{3}^{2}+.....
\end{equation}
For $n=1$, we regain the results for the fully annealed situation, given by
Eq. (\ref{l2}). From the coefficient of $I_{3}^{2}$, we immediately see that
the biaxial nematic phase is stable for $n<9/10$, which indicates that a
relatively weak departure from the fully annealed situation is already enough
to produce a stable biaxial structure.

In conclusion, we use a basic Maier-Saupe model of a binary mixture of
cylinders and discs to investigate the stability of a biaxial nematic phase.
In the quenched case, the biaxial phase is thermodynamically stable. In the
fully annealed situation, it becomes unstable, but in a situation of partial
annealing, represented by couplings to heat reservoirs at different
temperatures, we may recover thermodynamical stability.\bigskip

\textbf{Acknowledgements:} We acknowledge illuminating discussions with Andre
P. Vieira.

\end{document}